# Transient current responses of organic electrochemical transistors: Evaluating ion diffusion, chemical capacitance, and series elements

24 10 09

**Juan Bisquert*[1], Nir Tessler[2]**

[1]Instituto de Tecnología Química (Universitat Politècnica de València-Agencia Estatal Consejo Superior de Investigaciones Científicas), Av. dels Tarongers, 46022, València, Spain.

[2]Andrew & Erna Viterbi Department of Electrical and Computer Engineering, Technion, Haifa 32000, Israel

Corresponding author: jbisquer@itq.upv.es

**Abstract**

For the successful implementation of organic electrochemical transistors in neuromorphic computing, bioelectronics, and real-time sensing applications it is essential to understand the factors that influence device switching times. Here we describe a physical-electrochemical model of the transient response to a step of the gate voltage. The model incorporates (1) ion diffusion inside the channel that governs the electronic conductivity, (2) horizontal electron transport, and (3) the external elements (capacitance, ionic resistance) of the ion dynamics in the electrolyte. We find a general expression of two different time constants that determine the vertical insertion process in terms of the kinetic parameters, in addition to the electronic transit time. We highlight the central role of the chemical capacitance in determining the modulation of the lateral conductivity. The different types of response of the drain current are classified, and we discuss the significance for synaptic operation in neuromorphic circuits. The model is confirmed by detailed simulations that enable to visualize the different ions distributions and dynamics.

## 1. Introduction

Organic electrochemical transistors (OECTs) are presently under investigation for a variety of applications, including bioelectronics, logic circuit components, and neuromorphic devices.[1-7] In the OECT the channel is formed by a mixed ionic-electronic conductor.[8,9] The variable electronic conductivity is obtained by insertion and extraction of ions from an electrolyte and subsequent ion diffusion in the channel, while the compensating electronic carriers are established from drain and source contacts.[10-12]

OECT are excellent for translating chemical signals, such as ions or neurotransmitters, into electrical signals, as well as for accurately controlling stable conductance states to efficiently emulate computational tasks performed by biological synapses such as short-term depression (STD), short term potentiation (STP), and long-term potentiation (LTP).[13] However, fully capitalizing on OECTs requires a deeper comprehension of their fundamental transistor operation mechanisms, particularly regarding transistor switching behaviors, which play a pivotal role in the training phase of the neural networks. Controlling the device relaxation times is essential to increase the nonvolatility of neuromorphic transistor elements, by slowing down the ionic response, or inducing electrochemical reactions.

The switching properties have been investigated recently[14-24] and different conclusions have been obtained, regarding effects of the size of the cation, asymmetry of cation injection and extraction. The main approach to analyze the switching transient is the Bernards-Malliaras (BM) model,[2] that captures the coupling of cation insertion and the compensating electronic charge, by combining the electronic transient time across the channel, $\tau_e$, and the time $\tau_i$, associated to series connection of the ionic resistance of the electrolyte and the gate.[25] However, it is generally understood that the diffusion step of cations inside the channel is a dominant factor of the switching dynamics, and this was not included in BM.[26,27] The diffusion charging has been investigated for many years, starting with the conducting polymer films,[28-30] and also in the organic transistors.[31,32] Recently, asymmetric switching times have been observed upon charging and discharging, and they have been interpreted as lateral diffusion currents along the channel.[24,33,34]

In our previous work, we developed a general transition line model of OECT considering drift electronic transport and ionic injection and diffusion across the organic film, driven by the gate voltage.[26] The derived analytical model describes the dependence of drain current on gate bias in a time-transient situation, according to the measurement outlined in Fig. 1a. However, experimental patterns are more complex, due to the influence of capacitive and resistive elements in the electrolyte and its interfaces to the gate and channel. The impedance expression of the general model, recently derived,[35] including series impedances in the electrolyte, provides additional insight to the transient behaviour in more complex situations. Here we apply the theory of transients to show the types of decays that can be expected, their physical interpretation, and their evolution with the ratios of internal parameters of the system.

There are many general treatments of the dynamics of OECT using complete simulation programs, that describe well different aspect of the dynamical behaviour.[1,36-38] However, since the device has so many potential parameters and phenomena, the simulations are not so easy to interpret in comparison to experiment. The focus of the present work is different. Here we aim to obtain the key dynamical components that govern the transient response using some simplifying assumptions. This will enable us to use equivalent circuits model in parallel to the time transient physical equations. Then we can obtain profound insight as how different effects interact and produce observed responses in current transients. We will provide a detailed analysis of the simplifying assumptions and their limitations and discuss how different experimental techniques can be used in concert to clarify physical interpretations. We also compare the results of the simplified model with full physical simulation of OECT.

## 2. Transients in the ion diffusion model

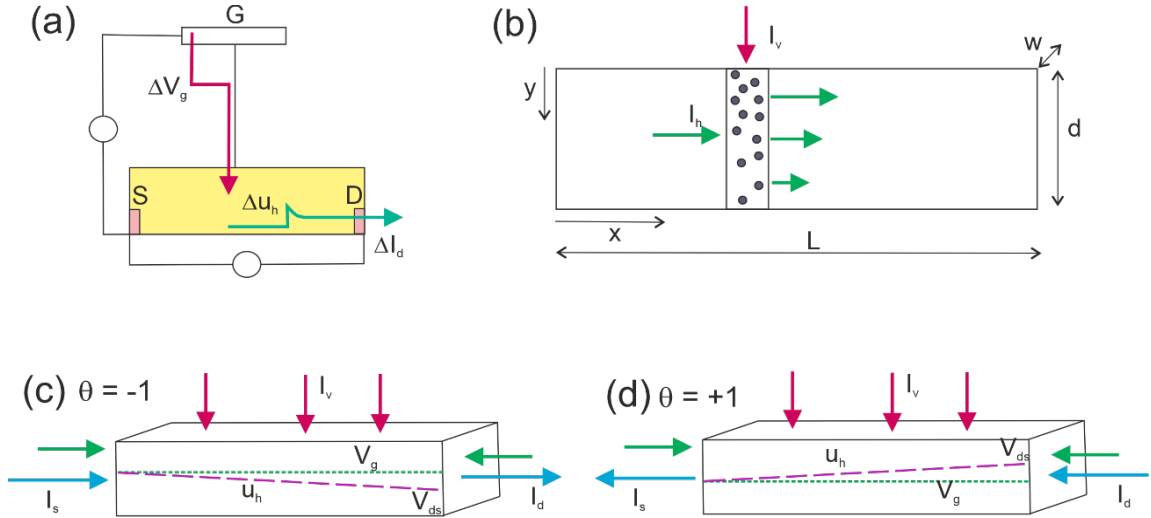

Fig. 1. (a) Scheme of the measurement of transient response of the OECT. (b) Scheme of distribution of currents. (c, d) The blue arrows indicate the stationary electronic current $I_d^{st} = I_s^{st}$. The green arrows are the transient charging electronic currents. In a transient situation, the difference of green currents at Drain and Source electrodes, $I_d - I_s$, equals the total ion current entering the channel film (red arrows). The voltage distribution for a small $u_{ds}$ is indicated for two situations: (c) $\theta = -1, V_{ds} < 0$, and (d) $\theta = +1, V_{ds} > 0$.

### 2.1. General transport-conservation equations

We summarize the previous results for the transients measured in the configuration of Fig. 1a. The main elements of the model are outlined in Fig. 1b.[26,35] The dominant physical processes consist of (a) ionic diffusion from the electrolyte to fill the channel at the equilibrium concentration imposed by the gate voltage (injected current $I_v$), combined with (b) drift electronic transport across the channel, of charge injected from source and drain contacts (horizontal current $I_h$). The general equations of the system have been described elsewhere for both accumulation and depletion mode OECT.[26,35] $L$ is channel length, $d$ is channel thickness and $w$ the width, $\mu_p$ is the hole mobility, and $D_{ion}$ is the ion diffusion coefficient inside the organic film.

The following conventions are used for the voltages and currents. The variable voltages are denoted $u_g$ (the gate voltage), $u_d$ (the drain voltage), etc.; the voltages of the stationary point are denoted $V_g, V_d$ ..., and finally, the small perturbation voltages are named $v_g, v_d$ ... The gate current is $I_g$, drain current is $I_d$, and the stationary currents are $I_k^{st}$. The $I_d$ is positive in the $x$ direction in Fig. 1b. Small perturbation currents are denoted $i_k$. The vertical small perturbation gate current is $i_g$ and the drain current is $i_d$.

Let us consider an accumulation mode OECT with injected anions of volume density $a$. The case of depletion can be treated similarly, as described before.[26] With reference to the simplified scheme of Fig. 1b, the vertical injection current depends on the diffusion flux $J_v$ at the boundary as

$$I_v = -q \, w \, J_v(y = 0) \tag{1}$$

and the charge conservation in a small volume implies that in the horizontal direction

$$\frac{dI_h}{dx} = I_v \tag{2}$$

On the other hand, the injected ions satisfy the diffusion equation and the conservation equation:

$$J_v = -D_{ion}\frac{\partial a}{\partial y} \tag{3}$$

$$\frac{\partial a}{\partial t} = -\frac{\partial J_v}{\partial y} \tag{4}$$

with the diffusion coefficient $D_{ion}$ and the boundary condition $J_v(y = d) = 0$.

The application of Kirchhoff rules combined with diffusion[29] produces the classical equations of a two coupled transmission line models that have been applied to OECT in many works.[18,37,39-41]

Under certain simplifying conditions of charge homogeneity in the vertical direction, that are discussed in Sec. 2.5, the vertical diffusion problem of Eqs. (3-4) can be reduced to the following two equations.[35]

$$I_g = C_\mu \frac{du_h}{dt} \tag{5}$$

$$I_g = -\frac{qL}{\tau_d}\left[A(u_g) - A(u_h)\right] \tag{6}$$

Here $q$ is the elementary charge, $I_g$ is the gate current, $u_h$ is the internal voltage in the channel, and $A$ is the average concentration of anions per unit horizontal distance in the channel. It is calculated from ions volume density $a(x, y, z)$ as

$$A = w \int_0^d a \, dy \tag{7}$$

The ion transport time in the finite layer is

$$\tau_d = \frac{d^2}{D_{ion}} \qquad (8)$$

The chemical capacitance is

$$C_\mu = -qL \frac{dA_{eq}}{du_g} \qquad (9)$$

For the horizontal current, Fig. 1c, d show the stationary (blue) and transient (green) source and drain current according to the direction of electrical field in the channel determined by the factor $\theta = V_{ds}/|V_{ds}|$ where $V_{ds}$ is drain-source voltage. Changing the sign of $\theta$ allows to turn the response of the drain into the corresponding response of the source current, which can also be measured. The current is described by the equation[26,35]

$$I_d = -\frac{q\theta L}{\tau_e} A - q L f_B \beta_\mu \frac{dA}{dt} \qquad (10)$$

here

$$\tau_e = \frac{L^2}{\mu_p |V_{ds}|} \qquad (11)$$

is the electronic transit time. The factor $0 < f_B < 1$ in Eq. (10) appears in the process of integration of Eq. (2). A transient electronic current compensating the vertical ionic current is attributed to the source, and the remaining fraction $f_B$ is attributed to the drain. This is explained in Ref. [26], and different methods to determine $f_B$ have been discussed in the literature.[18,20,42,43] The factor

$$\beta_\mu = \frac{1}{1 + \frac{A}{\mu_p}\frac{d\mu_p}{dA}} \qquad (12)$$

is due to the density dependence of the mobility.[44] Note that it can be $\beta_\mu < 0$ if the mobility $d\mu_p/dA < 0$, which a common property.[6,44,45]

In summary the model we use is composed of Eq. (10), that is already obtained in the BM model,[15] and Eq. (6), which is a simplified coupled diffusion equation, that is further discussed in Sec. 2.5.[26]

### 2.2. The transient response to gate voltage step

Now we take an operating point $V_g$ and we apply a small gate bias step perturbation $\Delta V$

$$v_g = \Delta V\, H(t) \qquad (13)$$

where $H(t)$ is the unit step function. Based on Eqs. (5, 6, 10) we can obtain the shape of the resulting current transients. We summarize the procedures explained in Ref. [26].

By solving Eqs. (5, 6) the time dependent voltage $v_h$ inside the channel is

$$v_h = \Delta V\left(1 - e^{-t/\tau_d}\right) \qquad (14)$$

We separate the drain current in Eq. (10) in two components: the stationary

$$I_{d0}^{st} = -\frac{q\theta L}{\tau_e}\, A_{eq}(V_g) \qquad (15)$$

and the time-dependent small perturbation component $i_d$

$$I_d(t) = I_{d0}^{st} + i_d(t) \qquad (16)$$

Eq. (10) gives the expression[26]

$$i_d(t) = C_\mu \left[ \frac{\theta}{\tau_e} v_h + \frac{f_B \beta_\mu}{\tau_d} \frac{dv_h}{dt} \right] \tag{17}$$

Using Eq. (14)

$$i_d(t) = C_\mu \left[ \frac{f_B \beta_\mu}{\tau_d} \Delta V + \left( \frac{\theta}{\beta_\mu \tau_e} - \frac{f_B \beta_\mu}{\tau_d} \right) v_h \right] \tag{18}$$

and we can write

$$i_d(t) = \left[ \frac{f_B \beta_\mu}{\tau_d} + \left( \frac{\theta}{\beta_\mu \tau_e} - \frac{f_B \beta_\mu}{\tau_d} \right) \left( 1 - e^{-t/\tau_d} \right) \right] C_\mu \Delta V \tag{19}$$

The vertical diffusion resistance[46] is

$$R_d = \frac{\tau_d}{C_\mu} \tag{20}$$

From Eq. (18), the initial current value is

$$i_{d0} = \frac{f_B \beta_\mu}{R_d} \Delta V \tag{21}$$

Therefore, the initial jump of the transient is a current through the vertical resistor $R_d$. This starts charging the channel film and increasing the charge density. More generally, the initial jump of the current depends on the total resistance in series with the capacitor, since $v_h(t = 0) = 0$, and the applied voltage goes to the series resistance.

The duration of the transient is set by $\tau_d$. The initial jump is independent of $\tau_e$. The final value of the step current is

$$i_{df} = C_\mu \frac{\theta}{\beta_\mu \tau_e} \Delta V \tag{22}$$

This value gives the post-jump equilibrium current, hence $I_{df}^{st} = I_{d0}^{st} + i_{df}$.

Using the initial and final currents, the current transient in Eq. (19) can be expressed as

$$i_d(t) = i_{df} \left[ 1 - \left( 1 - \theta \frac{\tau_h}{\tau_d} \right) e^{-t/\tau_d} \right] \tag{23}$$

In the case $\beta_\mu = \theta = 1$ we have

$$i_d(t) = i_{df} \left[ 1 - \left( 1 - \frac{f_B \tau_e}{\tau_d} \right) e^{-t/\tau_d} \right] \tag{24}$$

This is the standard expression of the transient in BM model,[15,20,24,41] but BM use an ionic time $\tau_i$ instead of the diffusion time $\tau_d$. We will explain this difference in Section 3.3.

To illustrate the transients in relation to a charging model in Fig. 2 we consider a concrete expression of the thermodynamic function[26]

$$A_{eq}(V_g) = \frac{2}{3} A_0 \left[ \frac{q(V_b - V_g)}{k_B T} \right]^{3/2} \tag{25}$$

with the corresponding chemical capacitance

$$C_\mu = -qL \frac{dA_{eq}}{d(V_g)} = \frac{A_0}{k_B T} \left[ \frac{q(V_b - V_g)}{k_B T} \right]^{1/2} \tag{26}$$

The $k_B$ is the Boltzmann's constant, $T$ the absolute temperature, $V_v$ the valence edge potential, $A_0$ a density per length.

Table 1

| channel length | $L$ | 50 μm |
|---|---|---|
| thickness | $d$ | 100 nm |
| width | $w$ | 10 μm |
| Hole mobility | $\mu_p$ | 0.02 cm²/Vs |
| Source-drain voltage | $|V_{ds}|$ | 0.1 V |
| Thermal voltage | $k_B T/q$ | 0.026 V |

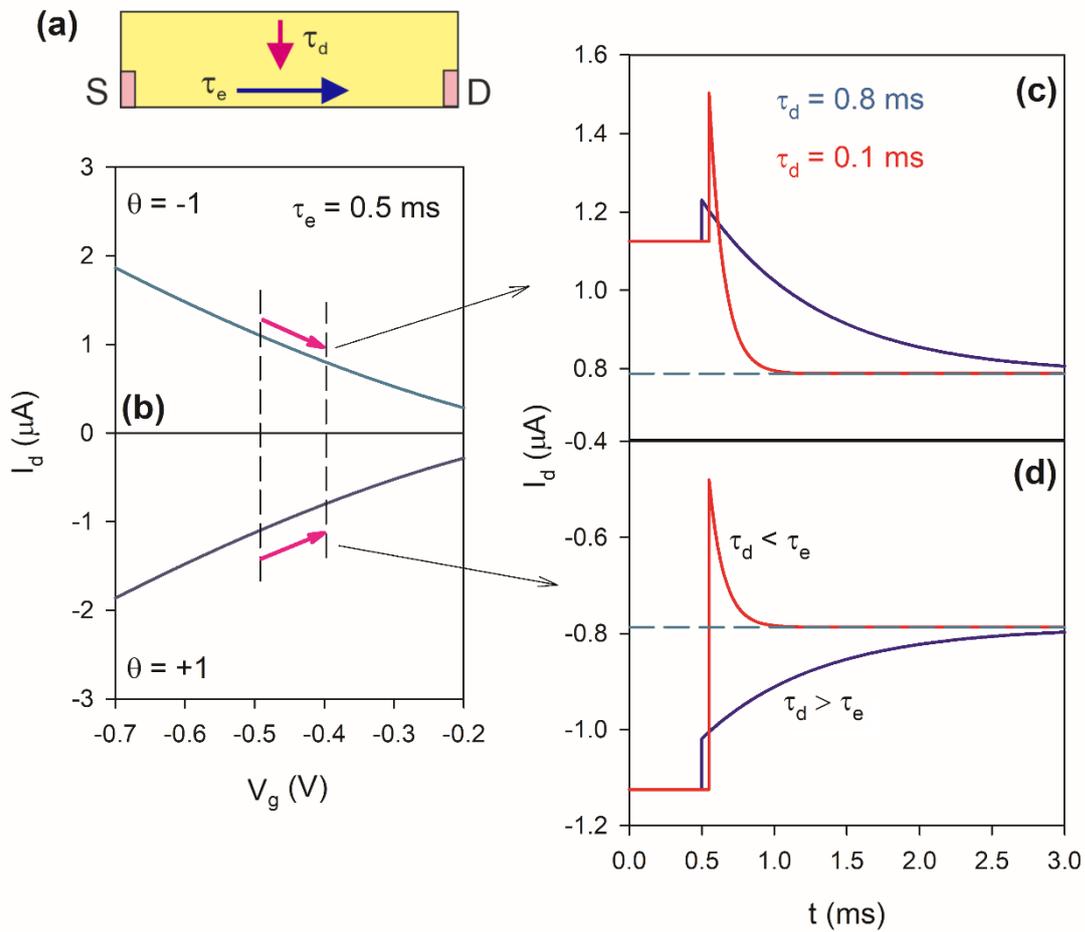

Fig. 2. (a) Scheme of the electrochemical transistor and the characteristic times processes. (b) The stationary current according to the sign $\theta = V_{ds}/|V|$ and (c, d) the possible four different types of transient response with respect to time, for a step voltage at $V_g = -0.5\,V$ and $\Delta V = 0.1\,V$ at $t_0 = 0.50, 0.51$ ms. $A_0 = 6.2 \times 10^{12}$ m⁻¹, $V_v = 0$, $f_B = 0.5$, rest of parameters in Table 1.

Fig. 2 shows four current transient conditions, according to the sign of $\theta$ and the relative size of the dominant relaxation times $\tau_e$ and $\tau_d$, by the parameters in Table 1. When $\theta = +1$, both jumps are larger than the final current. But if $\theta = -1$, one is larger and the other one smaller, so there is a point $\tau_e \approx \tau_d/f_B$ where the transient appears to vanish.

These transients are highly characteristic, as shown in Fig. 3[18] for a OECT made of the organic semiconductor poly(3,4-ethylenedioxythiophene) doped with poly(styrene sulfonate) (PEDOT:PSS). The PSS anions provide a counter charge for holes on the PEDOT chain and make the OECT channel conductive. A positive input voltage at the gate electrode modulates the channel current by pushing cations from the electrolyte into the PEDOT:PSS matrix. By changing the $V_{ds}$, one can modify the $\tau_e$, Eq. (11). Fig. 3 presents the three possible cases of transients according to Fig. 2d: Fig. 3a is for $\tau_e > \tau_d$; Fig. 3c is for $\tau_e < \tau_d$, and in Fig. 3b it is $\tau_e \approx \tau_d/f_B$ and the transient appears to disappear according to Eq. (24).

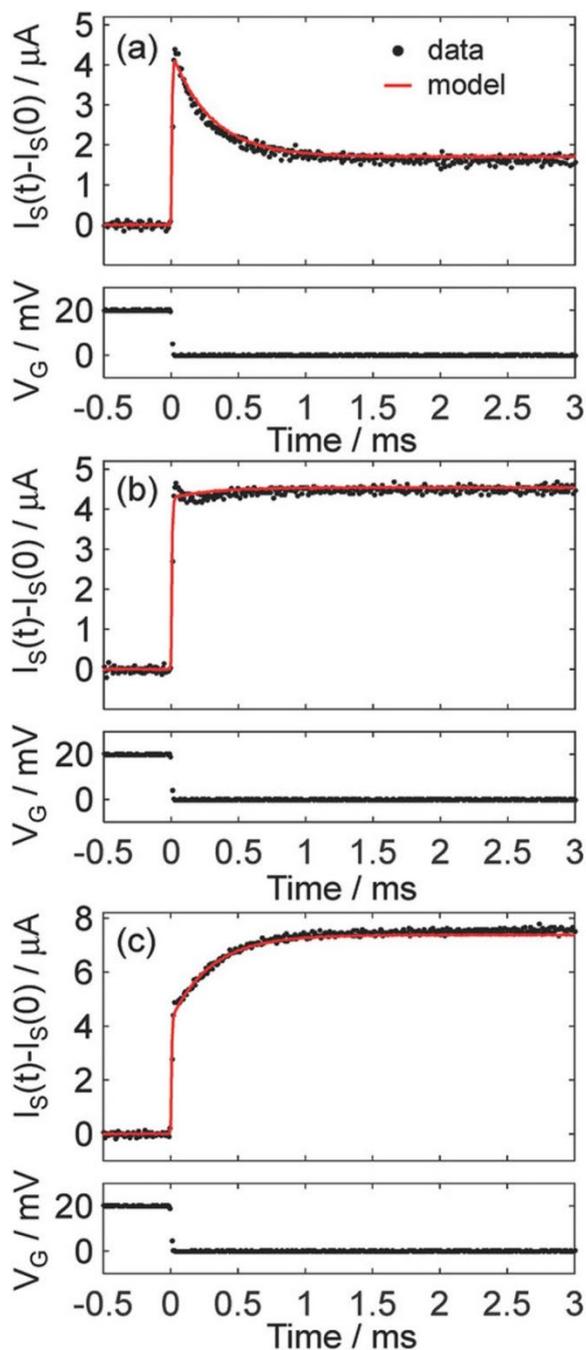

Fig. 3. Experimental results of a a–c) Change in the source current following a gate voltage step. The drain voltage is −30 mV for (a) −80 mV for (b) and −130 mV for (c) The gate voltage waveform is shown in the lower one-third of the panel in (a)–(c). The device had a channel width of W = 260 μm, a channel length of L = 100 μm, and a channel thickness of approximately h = 150 nm. Reproduced with permission from Friedlein, J.T., Donahue, M.J., Shaheen, S.E., Malliaras, G.G. and McLeod, R.R. (2016), Microsecond Response in Organic Electrochemical Transistors: Exceeding the Ionic Speed Limit. Adv. Mater., 28: 8398-8404. © 2016 WILEY-VCH Verlag GmbH & Co. KGaA, Weinheim[18].

Note that the shapes of the small perturbation transients do not depend on the specific thermodynamic model of ion and electronic carriers in the channel. The chemical capacitance only gives the size of the final current, Eq. (17). We can use any thermodynamic model that shows reasonable behaviour (i.e., it must satisfy $C_\mu > 0$, avoiding phase transformations). For example, the model

$$A_{eq}(V_g) = \frac{C^*}{qL}(V_0 - V_g) \tag{27}$$

gives a constant chemical capacitance that is amply used[10]

$$C_\mu = C^* \tag{28}$$

The shape of the transients will be the same as in the Fig. 2.

The linearized equations that are found in BM and in the above diffusion model provide significant insight to the physical components of the transient. However, for a large step of $V_g$ the linear model cannot be used, and it is necessary to solve the nonlinear system formed by Eqs. (5, 6, 10), or a more general simulation program, as in Sec. 5 of this paper. Due to nonlinear behaviour, on and off switching times can be different,[47] as observed experimentally in "asymmetric" switching times.[33]

### 2.3. The vertical impedance

To represent the small signal equations in the frequency domain, we make $d/dt \to s = i\omega$, where $\omega$ is the angular frequency and $i = \sqrt{-1}$. We have

$$i_g = \frac{1}{R_d}(v_g - v_h) \tag{29}$$

$$i_g = C_\mu s \, v_h \tag{30}$$

Combining Eqs. (29) and (30) we obtain

$$i_g = \frac{1}{Z_d} v_g \tag{31}$$

where the diffusion impedance is

$$Z_d = R_d + \frac{1}{C_\mu s} \tag{32}$$

The vertical impedance in Eq. (32) can be represented as an equivalent circuit shown in Fig. 4a.[35]

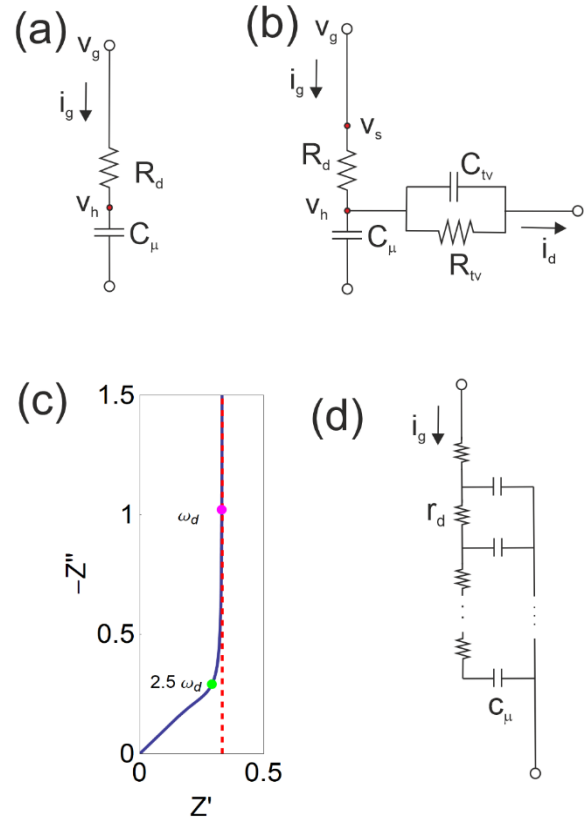

Fig. 4. (a) The equivalent circuit for the vertical small signal ac impedance. (b) The equivalent circuit for the transversal impedance. (c) Complex plot impedance spectra. The blue line is the finite transmission line model in (d), and the red line is the $R_d C_\mu$ model in (a). The points indicate characteristic frequencies, $\omega_d = 1/\tau_d$, and the frequency of the ankle, $\omega_g = (\pi^2/2)\,\omega_d$ (d) Transmission line diffusion impedance of a finite layer with blocking boundary condition.

If we do not adopt the approximation (6), the vertical diffusion problem in a finite layer of Eqs. (3, 4) produces the transmission line model diffusion impedance given by[46]

$$Z_{tl} = \frac{R_d}{(s/\omega_d)^{1/2}}\,Cotanh\big[(s/\omega_d)^{1/2}\big] \tag{33}$$

The model is shown in Fig. 4d, and the impedance spectra are shown in Fig. 4c. In Eq. (33)

$$\omega_d = \frac{1}{\tau_d} \tag{34}$$

is the characteristic frequency of diffusion. It separates roughly the domain of spatially infinite diffusion, that displays a 45º line, and the low frequency behavior, that is a vertical line. The full transmission line model is shown in blue line, and the $RC_\mu$ approximation, where the series resistance is $R = R_d/3$, is shown in the vertical red line. It is indicated also the angular frequency

$$\omega_g = \frac{\pi^2}{2}\frac{1}{\tau_d} \tag{35}$$

that marks the ankle of the transition between the two regimes. [46]

### 2.4. The transversal impedance

The horizontal current of Eq. (10) can be written

$$I_d = -\theta\frac{qL}{\tau_e}A + C_{tv}\frac{du_h}{dt} \tag{36}$$

where

$$C_{tv} = f_B\beta_\mu C_\mu \tag{37}$$

The transversal capacitance $C_{tv}$ in Eq. (37) is the chemical capacitance of holes in the channel. It is the same as the chemical capacitance of ions $C_\mu$ due to electroneutrality, but the factors in Eq. (37) ($f_B, \beta_\mu$) indicate the part of the total chemical capacitance that contributes to the drain current.

The modulation of $v_h$ induces a change of the drain current, according to the equation[35]

$$i_d = \left(\frac{1}{R_{tv}} + C_{tv}s\right)v_h \tag{38}$$

where the transversal resistance

$$R_{tv} = \frac{\theta\,\beta_\mu\,\tau_e}{C_\mu} = \frac{\theta\,\beta_\mu L^2}{C_\mu\mu_p|u_{ds}|} \tag{39}$$

is associated to the modulation of the electronic carrier density by the gate voltage.

The effective relaxation time of the electronic channel,

$$R_{tv}C_{tv} = \theta f_B\beta_\mu^2\tau_e = \theta\tau_h \tag{40}$$

where

$$\tau_h = f_B\beta_\mu^2\tau_e \tag{41}$$

The transverse impedance $v_g/i_d$ is represented as an equivalent circuit in Fig. 4b.[35]

### 2.5. Evaluation of the $RC$ approximation to diffusion

In the present and previous related works,[26,35] we have converted Eqs. (1-4) into a simplified model formed by Eqs. (6, 10). We call the Eq. (6) a simplified "diffusion" equation. By the transient solution of this system given in Eq. (14), one can use the Eq. (10), which is also a simplified solution of the horizontal transmission line problem in Eq. (2), and one obtains the transients of $i_d$ that we have described before in Fig. 2. We remark that also the BM model consists essentially of a $RC$ system, to which the horizontal transport Equation (10) is added.

To apply the mentioned approximations, certain conditions must be satisfied in the way that the ionic-electronic charge is distributed in the channel film. We can describe the situation with reference to Fig. 1b. When the dopant ions are inserted, they must produce a nearly homogeneous vertical distribution. Otherwise, the hole horizontal current depends on height $y$, and Eqs. (2-3) become two coupled transmission lines that can only be solved by the usual simulation programs.[18,37,39-41] If, on the other hand, Eq. (6) is accepted, then it can be combined with the BM-type approximation of Eq. (2), namely, Eq. (10), and valuable solutions are obtained.

The question is, when are these solutions reliable and valuable for describing experimental results?

Intuitively we can see that if the measurement time is much shorter than $\tau_d$, the charge distribution will be inhomogeneous, as the perturbation injected at the boundary with the electrolyte does not yet extend deeply to the channel hiegth. When $t > \tau_d$ it is likely that a homogeneous distribution is achieved, and the above approximations are valid.

To evaluate this picture quantitatively we need to solve the diffusion problem formed by Eqs. (3, 4), when a step concentration $a(0,t) = a_0$ is applied to a layer where initially $a = 0$ and $J_v(y = d) = 0$ at all times. The solution is given in Ref. [46]:

$$\frac{a(y,t)}{a_0} = 1 - 2\sum_{m=1}^{\infty} \beta_m^{-1} \sin\left(\beta_m \frac{y}{d}\right) \exp\left(-\beta_m^2 \frac{D_n t}{d^2}\right) \tag{42}$$

where $\beta_m = \pi\left(m - \frac{1}{2}\right)$ with $m = 1,2,\dots$

There are two important limit cases. At late times $t > \tau_d$, only the $m = 1$ term in the above sum contributes substantially, and $a(x,t)$ is well approximated by

$$\frac{a(y,t)}{a_0} = 1 - \frac{4}{\pi} \sin\left(\frac{\pi}{2}\frac{y}{d}\right) \exp\left(-\frac{t}{T_d}\right) \tag{43}$$

here

$$T_d = \frac{4}{\pi^2}\tau_d \tag{44}$$

Conversely, at early times, diffusion is hardly affected by the blocking boundary at $x = L$, and we may solve Eq. (4) in a semi-infinite geometry $y \in [0, \infty)$ instead, yielding

$$\frac{a(y,t)}{a_0} = \text{Erfc}\left(\frac{y}{\sqrt{4D_{ion}t}}\right) \tag{45}$$

in terms of the error function.

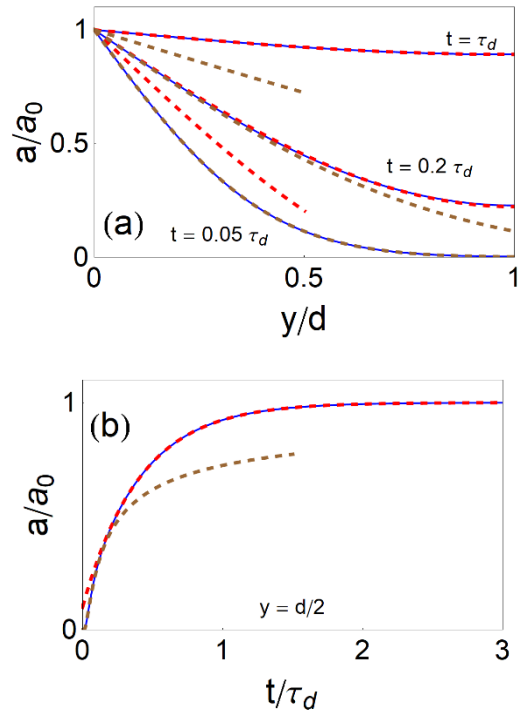

Fig. 5. The charging of a layer of thickness $d$ by application of a step of concentration $a_0$ at the left boundary. The blue line is the full solution, the red line is the exponential approximation, and the brown line is the Erfc approximation. (a) Carrier distribution into the layer at different times. (b) The concentration at $y = d/2$ as a function of time.

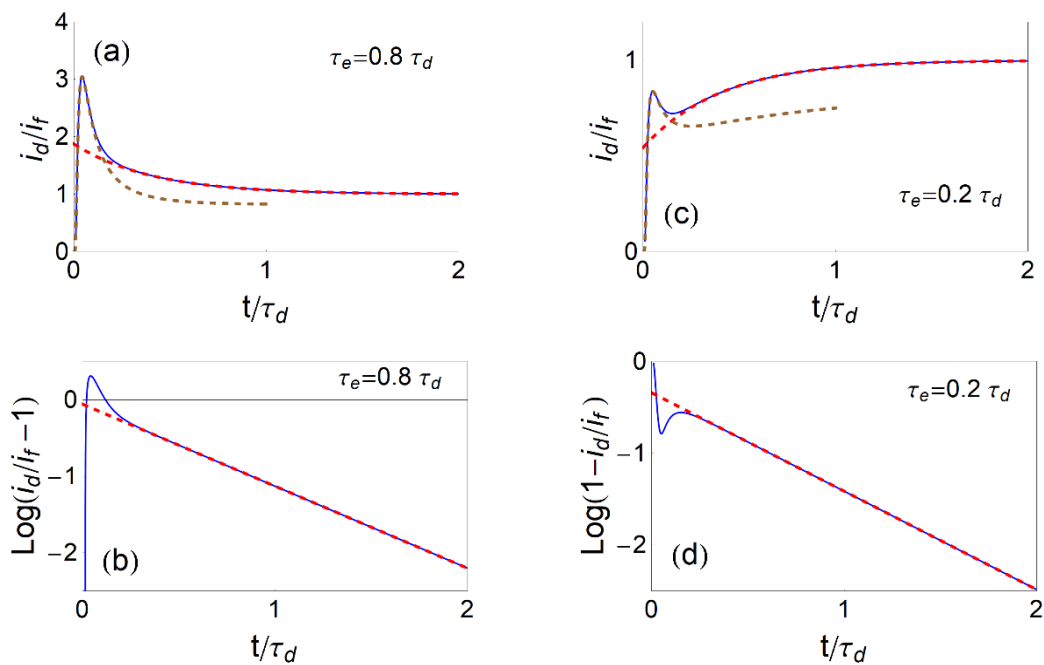

Fig. 6. Evolution of the drain current during the charging of the channel, calculated by Eq. (10) taking the concentration at $y = d/2$. The blue line is the full solution, the red line is the exponential approximation, and the brown line is the Erfc approximation. (a) and (c) are for different ratios of the characteristic times, the panels below show the same transient in Log vertical axis.

In Fig. 5a is represented the solution of charging the layer at different times (blue). At very short time the exponential approximation (red) is far from accurate, while the Erfc approximation describes the transient well. But at $t = 0.2\,\tau_d$ and longer time the exponential solution is quite accurate. In Fig. 5b is shown the concentration at the midpoint of the layer as a function of time. Again at $t = 0.2\,\tau_d$ and longer times the red exponential approximation is quite good.

In Fig. 6 is shown the source current obtained applying the Eq. (10) to the concentration at midpoint as a function of time, for the two cases of Fig. 2d. Since an average charge is used as $A$ in Eq. (10), we can see that in the very first instants this approximation cannot be applied, due to strong ions inhomogeneity in the vertical direction. But for $t = 0.2\,\tau_d$ and longer times the approximation of the exponential decay provides a good description of the excess transient current (for $\tau_e > \tau_d$, Fig. 6a, b) or the defect current (for $\tau_e < \tau_d$, Fig. 6c, d). In the first instants there is an excess or defect peak over the exponential line, that is due to the initial Erfc response.

Based on this analysis we obtain better insight to the significance of the approximations. With respect to the impedance response in Fig. 4c, the initial brown line of the decay in Fig. 6 corresponds to the 45º Warburg line. And the exponential red line of the transient corresponds to the vertical $RC$ line in Fig. 4c. By comparing the frequency of the ankle, $f_g = \omega_g/2\pi$, and the decay time $T_d$ in Eq. (43), we can see that they are related by

$$f_g = \frac{1}{\pi}\frac{1}{T_d} \tag{46}$$

We now state the validity of the developments based on the simplified diffusion equation (10), or the utilization of the $R_d C_\mu$ circuit instead of the transmission line. This approximation cannot describe the impedance spectra or the current transients at frequencies higher or times shorter than indicated in Eq. (46). Such "fast" domains cannot be treated by the methods described in this paper; they must be treated separately. But at the longer times, the simplified analysis is quite useful. In the measurement, a small overshoot will appear in the log representation as shown in Fig. 6b. This is the initial diffusion domain. Thereafter, the exponential approximation is quite good, and enables to combine the diffusion charging with other aspects of the system, as we discuss in the Section 3 of this paper.

### 2.6. Depletion mode transistor

This is the case $A = 0$, cation density $Z = M$, and hole carrier density $P = P_0 - M$, where $P_0$ is the initial doping. The model equations are[26]

$$I_h(L) = -\theta \frac{qL}{\tau_e}(P_0 - M) + qfL\frac{dM}{dt} \qquad (47)$$

$$\tau_d \frac{dM}{dt} = M_{eq}(u_g) - M \qquad (48)$$

In the small perturbation we obtain

$$I_d = -\theta \frac{qL}{\tau_e}\left(P_0 - M(u_{g0})\right) + C_\mu\left(\frac{\theta}{\tau_e}v_h + f_B\frac{dv_h}{dt}\right) \qquad (49)$$

Comparing with Eq. (17), we remark that the stationary expressions are different in accumulation and depletion, but the small perturbation method can be applied equally to both. Hence the forthcoming extensions apply equally to both types.

## 3. Model of transients with diffusion and interfacial/electrolyte impedances

### 3.1. The vertical current/voltage equations

To represent the electrolyte and interfacial limitations we add to the model of Fig. 4a the parallel combination of ionic resistance $R_i$ and double layer capacitance $C_d$ as shown in Fig. 7a. We aim to find a solution of the transient voltage and current for the extended system. The new vertical circuit is represented in the equivalent circuit of Fig. 7b. We remark that in an OCET model the $C_d$ must be combined with a *parallel* resistance, since the ionic current must penetrate the channel, and cannot be blocked at the interface.

The vertical current provides three equations[26]

$$I_g = \frac{1}{R_i}(u_g - u_s) + C_d\frac{d(u_g - u_s)}{dt} \qquad (50)$$

$$I_g = -\frac{qL}{\tau_d}[A(u_s) - A(u_h)] \qquad (51)$$

$$I_g = C_\mu\frac{du_h}{dt} \qquad (52)$$

Here $u_s$ is the voltage at the outer surface of the channel.

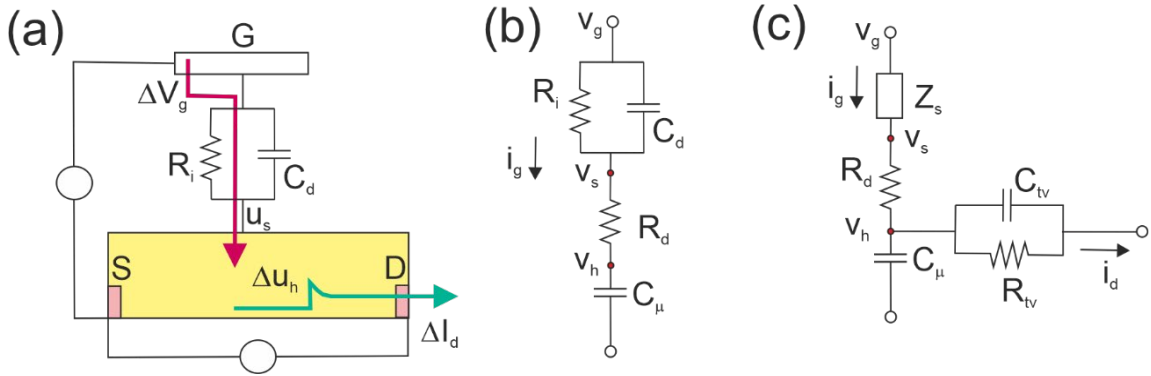

Fig. 7. (a) Scheme of the measurement of transient response of the OECT with series resistance and capacitor in the electrolyte. (b) The equivalent circuit for the vertical small signal ac impedance. (c) The equivalent circuit for the transversal impedance.

### 3.2. The transient of the vertical current

To obtain the time-dependent currents in Fig. 7a, we apply the same small perturbation method as in Sec. 2. We solve the response to an applied signal in the frequency domain. The solution in the time domain will be then obtained by inverse Laplace transform.

The vertical current equations become

$$i_g = \left( \frac{1}{R_i} + C_d s \right) \left( v_g - v_s \right) \tag{53}$$

$$i_g = \frac{1}{R_d} \left( v_s - v_h \right) \tag{54}$$

$$i_g = C_\mu v_h \tag{55}$$

We define the relaxation times

$$\tau_s = R_i C_d \tag{56}$$

$$\tau_{R\mu} = R_i C_\mu \tag{57}$$

The above equations yield the relation

$$v_h = \frac{1 + \tau_s s}{(1 + \tau_s s)(1 + \tau_d s) + \tau_{R\mu} s} v_g \tag{58}$$

Eq. (58) is the response of the internal voltage in the channel to the gate voltage perturbation.

Solving the quadratic equation in the denominator, and noting that $v_g = \Delta V / s$, we can write

$$v_h = \frac{1 + \tau_s s}{s(1 + \tau_1 s)(1 + \tau_2 s)} \Delta V \tag{59}$$

The fundamental relaxation times are

$$\tau_1 = 2 \frac{\tau_0}{1 - b} \tag{60}$$

$$\tau_2 = 2 \frac{\tau_0}{1 + b} \tag{61}$$

where

$$\tau_0 = \left[ \frac{1}{\tau_d} \left( 1 + \frac{C_\mu}{C_d} \right) + \frac{1}{\tau_s} \right]^{-1} \tag{62}$$

$$b = \left( 1 - 4 \frac{\tau_0^2}{\tau_d \tau_s} \right)^{1/2} \tag{63}$$

By inversion of (59) to the time domain we have the solution

$$v_h = \left[ 1 - \left( \delta_1 e^{-t/\tau_1} + \delta_2 e^{-t/\tau_2} \right) \right] \Delta V \tag{64}$$

$$\delta_1 = \frac{\tau_0}{b \, \tau_d} \left( \frac{\tau_1}{\tau_s} - 1 \right) \tag{65}$$

$$\delta_2 = \frac{\tau_0}{b \, \tau_d} \left( 1 - \frac{\tau_2}{\tau_s} \right) \tag{66}$$

Eq. (64) is the relaxation of the internal voltage when a pulse $v_g = \Delta V$ is applied to the gate contact. This result extends Eq. (17) to the more general situation. Note that $v_h(0) = 0$, $v_h(\infty) = \Delta V$. The $v_h(t)$ is not dependent on $\tau_e$ since there is no horizontal component of the perturbation. However, this simplification will change if $\tau_e$ is very long and influences the diffusion transport.[26]

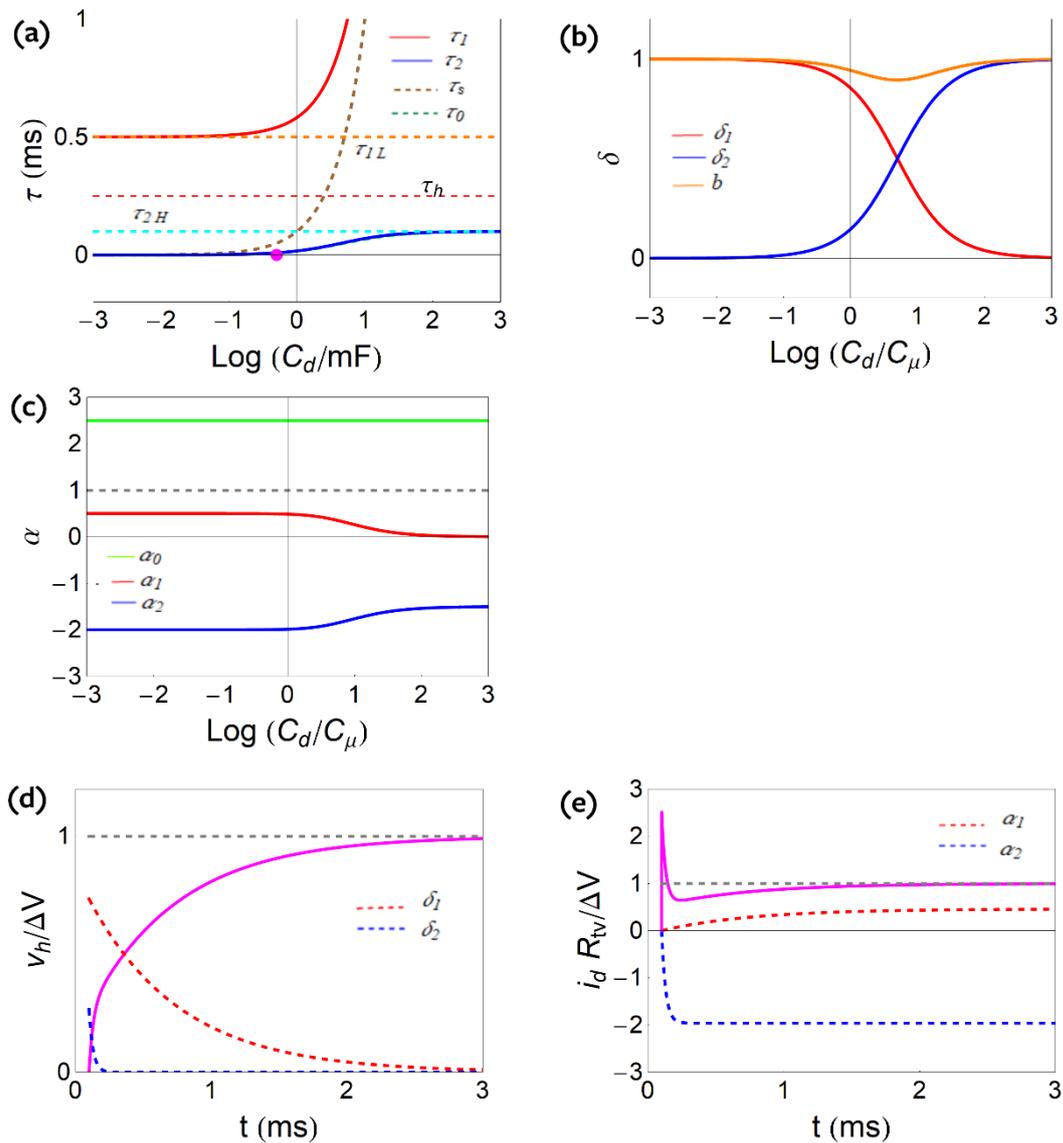

Fig. 8. Model simulations of the system of Fig. 7. (a) The relaxation times. The point indicates the value used in (d, e). (b, c) The weight factors of the transients. Parameters: $\tau_d = 0.1\ ms$, $\tau_e = 0.5\ ms$, $f_B = 0.5$, $R_i = 0.2\ \Omega$, $C_\mu = 2\ mF$. (d, e) The transients of $v_h$ and $i_d$ for $C_d = 1\ mF$. The grey dashed line is the final equilibrium.

For the interpretation of these results, we turn to Fig. 8. The two fundamental decay time constants, $\tau_1$ and $\tau_2$ are plotted with respect to $C_d/C_\mu$ in Fig. 8a, and the weights in the voltage transient $\delta_i$ are shown in Fig. 8b. Note that $\tau_2 \approx \tau_0$ at all values of $C_d$. The voltage transient is shown in Fig. 8d, indicating the contribution of each component, for a particular value of $C_d$ (the dot in Fig. 8a).

The times $\tau_1$ and $\tau_2$ become independent of $C_d$ at low and high $C_d$, respectively. To obtain the limiting values we first find the limits of $\tau_0$. For $C_d \ll C_\mu$ it is

$$\tau_{0L} = \left[\frac{1}{C_d R_d} + \frac{1}{\tau_s}\right]^{-1} = \left[\frac{1}{R_d} + \frac{1}{R_i}\right]^{-1} C_d \tag{67}$$

This is the charging of the surface capacitor $C_d$. While for $C_d \gg C_\mu$

$$\tau_{0H} = \left[\frac{1}{\tau_d} + \frac{1}{\tau_s}\right]^{-1} \tag{68}$$

Here is the charging of the surface capacitor limited by the charging of the chemical capacitance when $C_d$ becomes large.

In Fig. 8b we note that for $C_d \ll C_\mu$ the decay is dominated by $\tau_1$, and it can be approximated as

$$\tau_{1L} = (R_i + R_d)C_\mu \tag{69}$$

For $C_d \gg C_\mu$, it is $\delta_1 \approx 0$ and $\delta_2 \approx 1$. Hence the second root dominates the decay for $C_d \gg C_\mu$, and it becomes

$$\tau_{2H} = \tau_{0H} = R_d C_\mu = \tau_d \tag{70}$$

Accordingly, we find that $\tau_2$ corresponds to the original time constant $\tau_d$ in the diffusion model of Fig. 1 and 2, while $\tau_1$ contains a contribution of the series resistance, Eq. (69), and it is always

$$\tau_{2H} < \tau_{1L} \tag{71}$$

and, more generally

$$\tau_2 < \tau_1 \tag{72}$$

Therefore, there are two components in the transient of $v_h$ towards $\Delta V$. $\tau_2$ is the fast component and $\tau_1$ the slow one, Fig. 8d. The diffusion charging ($\tau_2$) happens first, and then occurs a slower charging due to the external resistor and capacitor ($\tau_1$).

### 3.3. The transient of the horizontal current

We can write Eq. (36) as

$$i_d = \frac{1}{R_{el}}(1 + \theta\tau_h s)v_h \tag{73}$$

Therefore

$$i_d = \frac{\theta}{R_{el}} \frac{(1+\tau_s s)(1+\theta\tau_h s)}{(1+\tau_1 s)(1+\tau_2 s)} \frac{\Delta V}{s} \tag{74}$$

This model corresponds to Fig. 7c.[35] The solution in the time domain is

$$i_d(t) = \left[\alpha_0 + \alpha_1\left(1 - e^{-\frac{t}{\tau_1}}\right) + \alpha_2\left(1 - e^{-\frac{t}{\tau_2}}\right)\right]\frac{\theta\,C_\mu}{\beta_\mu\tau_e}\Delta V \tag{75}$$

where

$$\alpha_0 = \frac{\theta \tau_s \tau_h}{\tau_1 \tau_2} \tag{76}$$

$$\alpha_1 = \frac{1}{(\tau_2 - \tau_1)} \left( -\tau_1 + \theta \tau_h + \tau_s - \frac{\theta \tau_h \tau_s}{\tau_1} \right) \tag{77}$$

$$\alpha_2 = \frac{1}{(\tau_2 - \tau_1)} \left( \tau_2 - \theta \tau_h - \tau_s + \frac{\theta \tau_h \tau_s}{\tau_2} \right) \tag{78}$$

We observe in Eq. (75) that the current transient contains two components governed by the fundamental time constants, $\tau_1$ and $\tau_2$. The weights of the components are shown in Fig. 8c. The initial value of the current response is

$$i_{d,in} = \alpha_0 \frac{\Delta V}{R_{tv}} \tag{79}$$

and the final value is

$$i_{d,fin} = \frac{\Delta V}{R_{el}} \tag{80}$$

As commented before, $\tau_2$ is the fast component, and this is observed in Fig. 8e, where the initial spike is the same as that in Fig. 2d, since $\tau_d < \tau_h$ in this example. But then the signal rises to a larger final value, due to the longer time constant $\tau_1$.

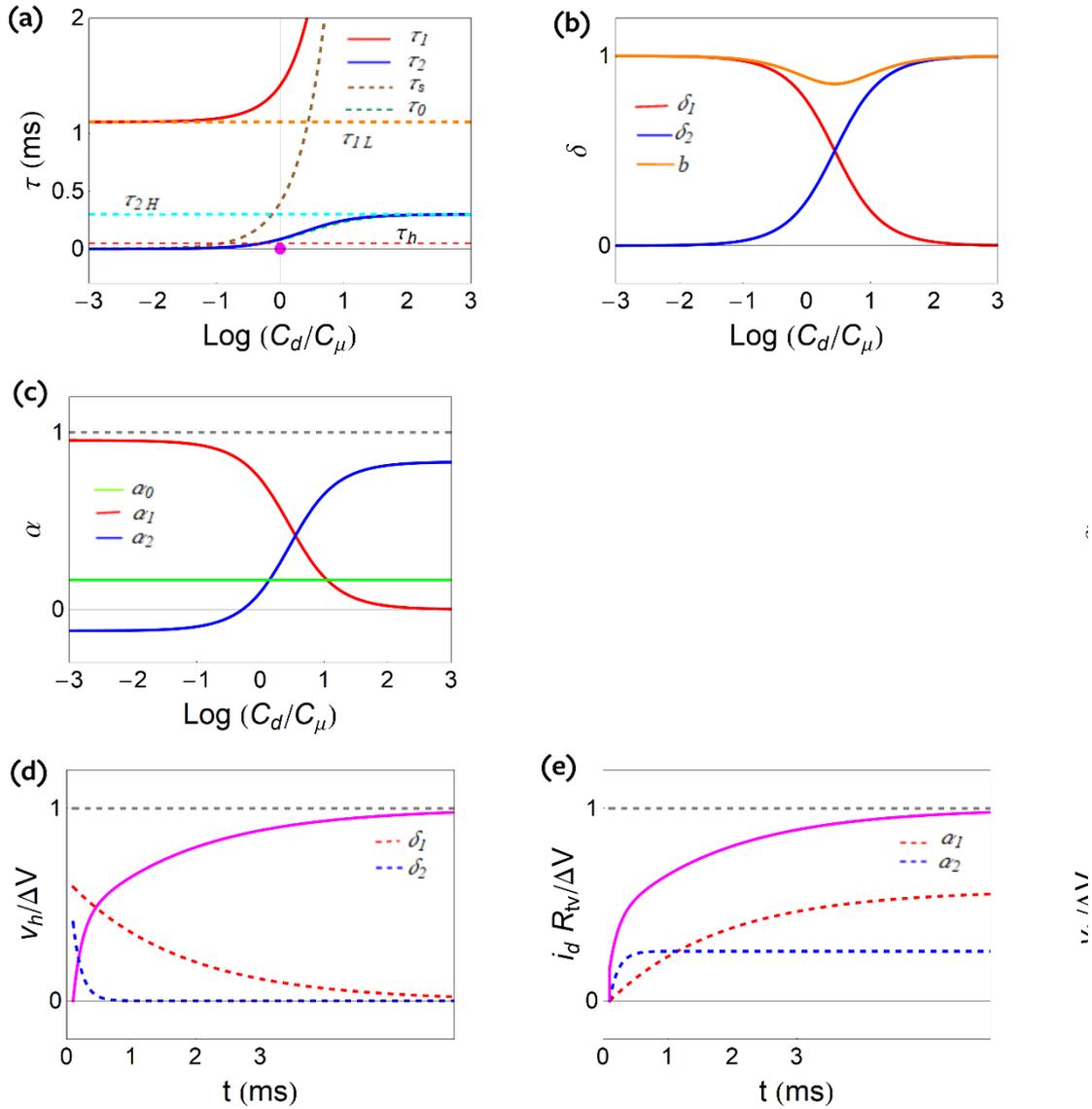

Fig. 9. Model simulations of the system of Fig. 7. (a) The relaxation times. The point indicates the value used in (d, e). (b, c) The weight factors of the transients. Parameters: $\tau_d = 0.3 \; ms, \tau_e = 0.1 \; ms, f_B = 0.5, R_i = 0.4 \; \Omega, \; C_\mu = 2 \; mF$. (d, e) The transients of $v_h$ and $i_d$ for $C_d = 2 \; mF$. The grey dashed line is the final equilibrium.

In Fig. 9 we show results for the opposite case in Fig. 2, $\tau_d > \tau_h$. In Fig. 9d we remark the rise of the voltage by the combination of the two fundamental time constants. In Fig. 9e we observe the components of the rise of the current: It starts at $\alpha_0 \approx 0.2$, shows a rapid rise by $\tau_2$, and a longer rise by $\tau_1$.

By Eq. (69) we can observe that

$$\tau_{1L} = R_i C_\mu \tag{81}$$

in the case in which the outer resistance is dominant. This is the time called $\tau_i$ in the Bernards-Malliaras model, where the transient is usually written as Eq. (24), with $\tau_i$ instead of the diffusion time $\tau_d$. Note that BM model obtains a $RC$ circuit similar to Fig.

4a but the resistance is interpreted as electrolyte transport.[15,17] Instead, in Fig. 4a $R_d$ and $C_\mu$ are associated to ion diffusion and accumulation inside the channel.[35]

We arrive at the conclusion that BM expression is a particular case of Eq. (69), that contains both the external ionic resistance and diffusion inside the channel, represented by a "volume capacitance".

Again, we remark that for a large voltage step the full set of nonlinear equations must be solved to obtain the transient response.

## 4. Interpretation of the transients

Based on the insights obtained in the preceding sections we provide more detailed interpretation for analysis of the transients of OECT and the identification of their components in experimental results.

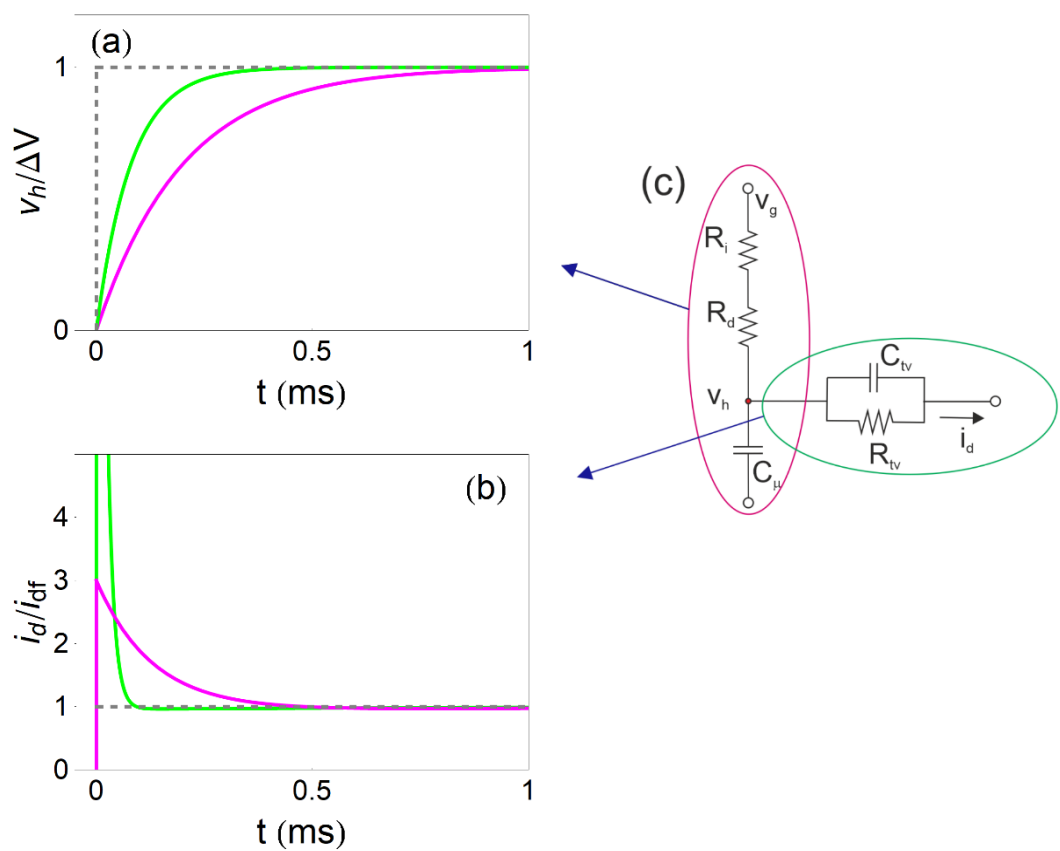

Fig. 10. The transients in a system with diffusion charging, without series elements ($R_i = 0\ \Omega$). (a) Voltage in the film, (b) drain current. Purple: $\tau_d = 0.2\ ms$, $R_d = 0.1\ \Omega$, green: $\tau_d = 0.08\ ms$, $R_d = 0.04\ \Omega$. Parameters: $\tau_e = 1.2\ ms$, $f_B = 0.5$, $C_\mu = 2\ mF$. (c) The equivalent circuit part that controls each transient. The transient of the voltage is determined by the vertical impedance (red). The transient of the drain current incorporates the horizontal channel impedance (green), to form the final transverse impedance.

As we commented in Fig. 1a the basic measurement consists of application of a step of the gate voltage and the measurement of the drain current. Consider in Fig. 10 a system in the absence of series elements, that was already solved in Fig. 2. There are two components to the transient. One is the exponential evolution of the internal voltage $v_h$, indicated in Eq. (17) and shown in Fig. 10a. This is associated fully to the vertical impedance. Next, the change of $v_h$ produces a change of $i_d$, given in Eq. (17), and the corresponding examples in Fig. 2b, shown also in Fig. 10b.

The exponential form in Eq. (17) arises from the $R_d C_\mu$ approximation that comes from Eq. (6). This is not accurate at times shorter than $\tau_d$, as we have discussed in Figs. 5 and 6. The $R_d C_\mu$ approximation cannot capture the initial diffusion part that corresponds to the 45º line in the impedance of Fig. 4c. Therefore, it is highly recommended to combine the analysis of transients with the measurement of impedance spectroscopy, that reveals the semi-infinite diffusion domain. Another important reason is that the $R_d C_\mu$ approximation cannot distinguish a possible series resistance $R_i$ from the true diffusion resistance $R_d$, as they are connected in series. This is indicated in Fig. 10c. The impedance spectra can produce such distinction, as a series resistance simply displaces the transmission line spectra.

Now we highlight the essential role of the chemical capacitance in the transient dynamics. By using the $R_d C_\mu$ approximation in Fig. 10c we observe that the chemical capacitance is the element that controls the conductivity of the channel and governs the interesting properties such as synapsis potentiation, as the charge in $C_\mu$ keeps the memory of previous stimuli. Here the $R_d C_\mu$ approximation provides a strong insight into the transistor operation, that is obscured if we keep the whole transmission line in the vertical impedance.

In both systems of Fig. 10a the chemical capacitance is the same. In the green system the ion diffusion coefficient is larger, the diffusion resistance is smaller, and thus the $\tau_d$ is smaller, than in the purple system. This difference translates into the different spikes of Fig. 10b, and the different forms of Fig. 2, according to the relation between $\tau_d$ and $\tau_e$.

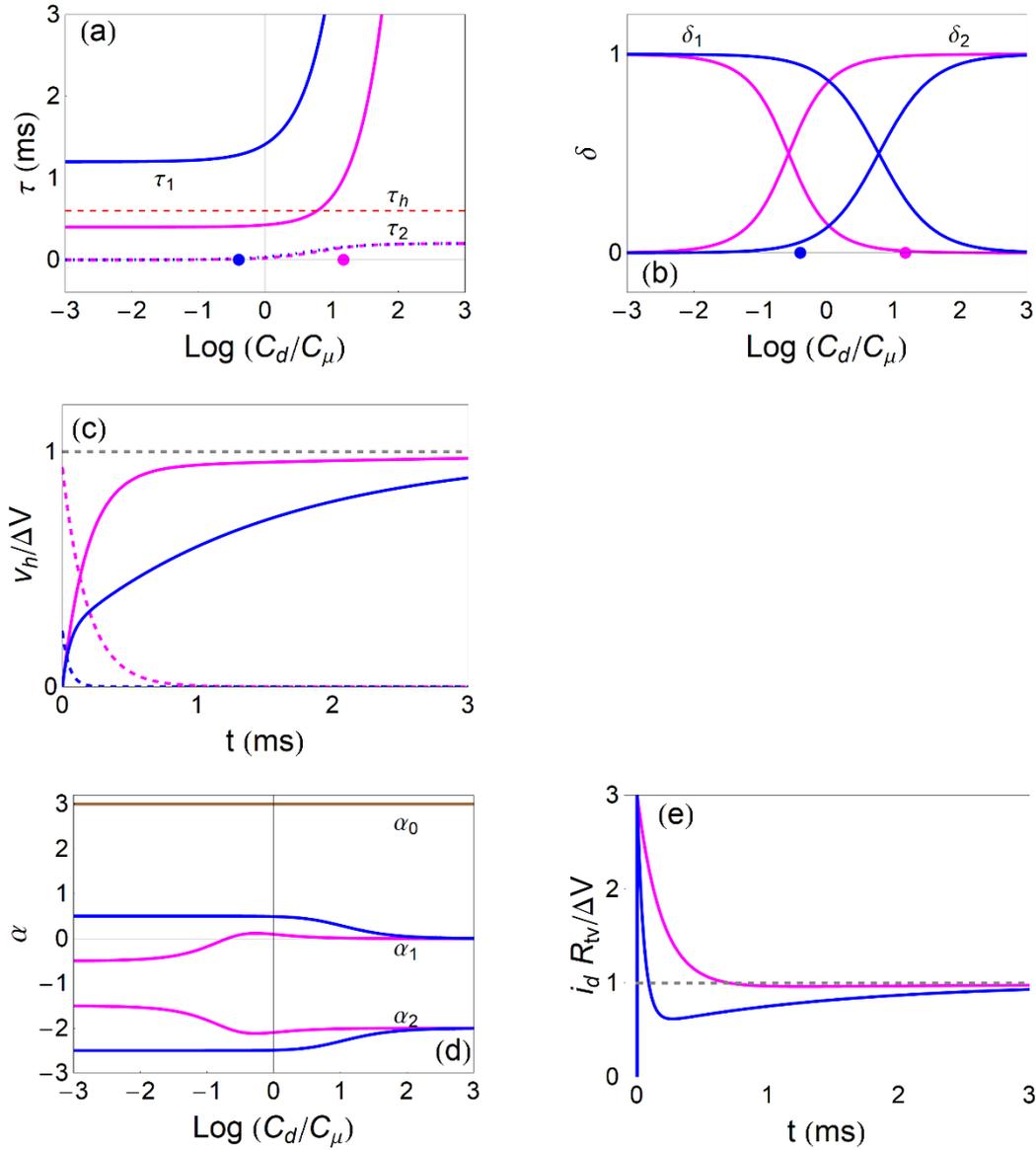

Fig. 11. Comparison of transients for the system of Fig. 7. Common parameters: Parameters: $\tau_d = 0.2\ ms$, $\tau_e = 1.2\ ms$, $f_B = 0.5$, $C_\mu = 2\ mF$, $R_d = 0.1\ \Omega$. Purple: $R_i = 0.1\ \Omega$, $C_d = 30\ mF$, blue: $R_i = 0.5\ \Omega$, $C_d = 0.8\ mF$. (a) Shows the relaxation times as function of $C_d/C_\mu$. The points in the x-axis indicate the values of $C_d$. (b) The weights of the voltage decay. (c) The evolution of $v_h$. The dashed lines show the decay of $\delta_2 e^{-t/\tau_2}$. (d) The amplitudes of the decay of $i_d$ as function of $C_d/C_\mu$. (e) The evolution of $i_d$.

As we have already mentioned, the chemical capacitance is a necessary part of the system, which gives the changes of carrier number in response to the change of gate voltage. But there are other possible capacitances in the system, as shown in Fig. 7. In Sec. 3 we have developed a long analysis to clarify their consequences in the dynamic response. We now present some additional cases of the system of Fig. 7 for discussion.

In Fig. 11 we compare two systems with the same chemical capacitance $C_\mu$. In the blue system $C_d = 0.8 \ mF$, and we see in Fig. 11b that the voltage transient is mainly controlled by $\tau_1$, while in the purple system $C_d = 30 \ mF$ and the transient is governed by $\tau_2$. Hence the voltage transient of the purple system is faster in Fig. 11c.

Another important feature is the different signatures of the drain current to approach equilibrium, either by direct decay, as in Fig. 2, or by an initial spike followed by a rise of current, that did not occur in Fig. 2. We find this composed pattern in Fig. 8e and in the blue line in Fig. 11e. This behaviour is relevant for synapsis operation. In two-contact devices we can distinguish two main patterns> decay, in capacitive systems, and a rise of current, in inductive systems, that are associated with synapse depression and potentiation, respectively.[47,48]

As shown in Fig. 2d in the transistor system the rise or decay behaviour is due to the relation between $\tau_d$ and $\tau_h$. We note in Eqs. (77, 78) that $\alpha_1$, that determines the long-time decay of $i_d$, depends on the difference of $\tau_1$ and $\tau_h$, while $\alpha_2$, that governs the short-time decay, depends on the difference of $\tau_2$ and $\tau_h$. In the magenta case of Fig. 11, $\tau_1 < \tau_h$. Since $\alpha_1$ is mostly negative, the decay of the current is "capacitive" and decreases monotonously. In the blue case $\tau_1 > \tau_h$, with $\alpha_1 > 0$, and the long-time current increases in "inductive" case. Thus $\tau_1$ in Fig. 11 plays the role of $\tau_d$ in Fig. 2d. But now $\tau_1$ is mainly due to the external $RC$ elements, not the diffusion charging. In addition, $\tau_2$ contains the fastest component of the decay, mainly determine by the diffusion insertion of ions. To confirm this interpretation, we show in Fig. 12 two more cases where the diffusion time $\tau_d$ is fully negligible with the respect to the ionic charging time (hence no dashed line is seen in Fig. 12b). Again, we can obtain in Fig. 12c the two cases of Fig. 2d, raising or decaying.

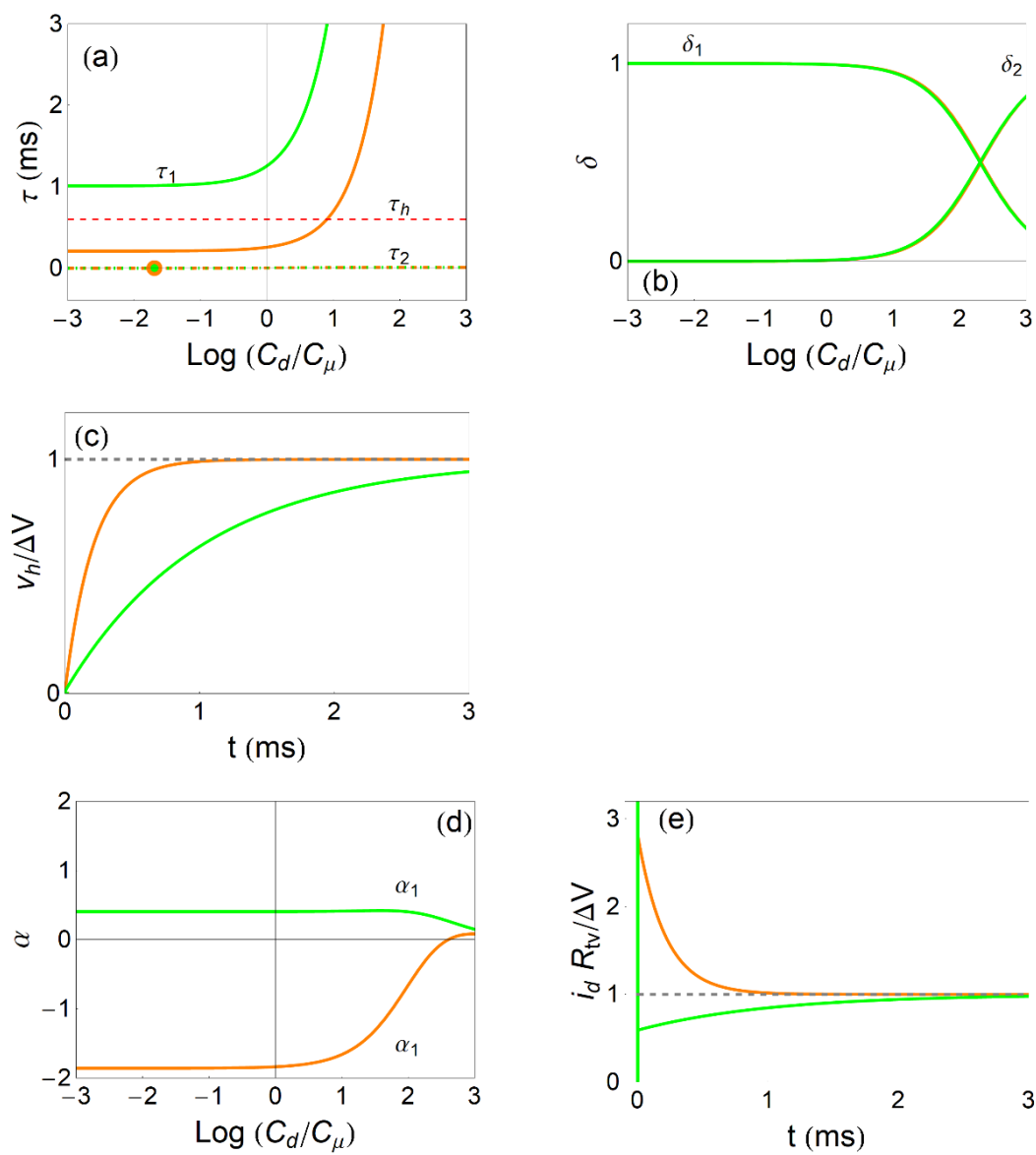

Fig. 12. Comparison of transients for the system of Fig. 7. Common parameters: Parameters: $\tau_d = 0.01\ ms, \tau_e = 1.2\ ms, f_B = 0.5,\ C_\mu = 2\ mF, R_d = 0.005\ \Omega$. Orange: $R_i = 0.1\ \Omega, C_d = 0.02\ mF$, purple: $R_i = 0.5\ \Omega, C_d = 0.02\ mF$. (a) Shows the relaxation times. The points in the x-axis indicate the values of $C_d$. (b) The weights of the voltage decay. (c) The evolution of $v_h$. (d) The amplitudes $\alpha_1$ of the decay of $i_d$ as function of $C_d/C_\mu$. ($\alpha_2 \approx \alpha_0 \approx -60$.) (e) The evolution of $i_d$.

## 5. Visualising the interface phenomena through 2D device modelling

The solubility difference is the most notable chemical-physics mechanism that defines the discontinuity associated with the interface between the solution and the semiconductor. This section aims to show how this would lead to an effective circuit of resistance in parallel to capacitance and that the resulting effect on the current transient agrees with the analytical model. We use the same simulation approach reported in refs. [23,26], which consists of employing the Sentaurus device TCAD by Synopsys to solve ionic and electronic transport within a semiconductor device model framework. To our previous reports,[23,26] we need to add the potential solubility difference at the solution/semiconductor interface with the ions being less soluble in the semiconductor. Henry's law for solubility[49] can be used to show that solubility difference dictates a concentration ratio between the two sides of the interface. Such a concentration (density) ratio corresponds to an energy barrier within a semiconductor device model framework. Hence, we will implement an energy barrier ($E_b$) that will enforce a density ratio of $exp(-E_b/k_BT)$.

Fig. 13a shows the OECT device structure used in the simulations, and Table 2 lists the parameters implemented in the simulations. It is a P type OECT based on an undoped semiconductor where the penetration of anions from the solution would dope and switch it on. We show the hole density distribution inside the semiconductor for $V_{DS}$=-0.1 V, $V_{GS}$=-0.05 V, and $I_{DS}$=10 μA (see red circle in Fig. 13b).

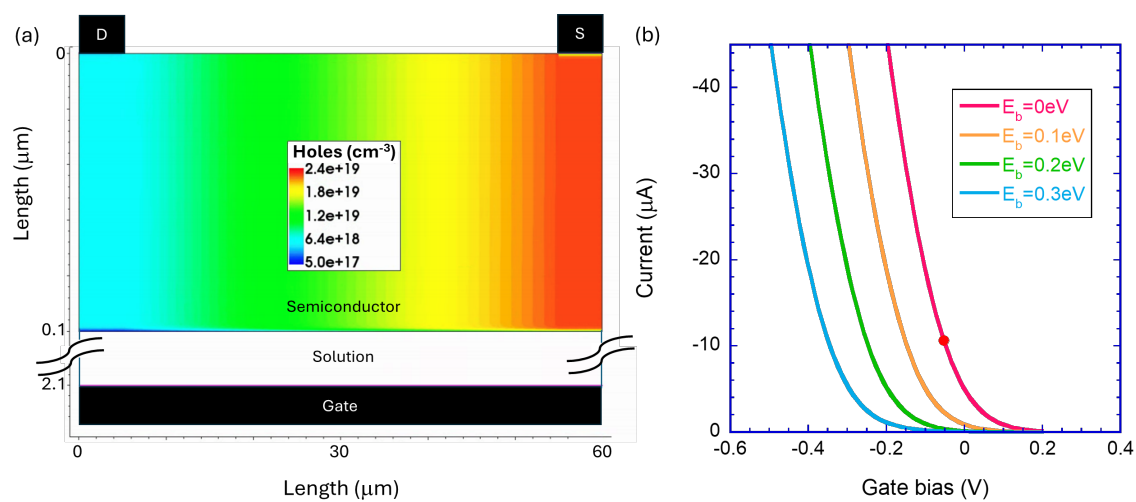

Fig. 13. (a) Schematic description of the OECT device structure exhibiting the hole density distribution for $V_{GS}$ = 0 V, $V_{DS}$ = -0.1 V, and no ionic injection barrier at the solution/semiconductor interface. (b) Current gate-voltage (output characteristics) of OECT devices for different ion injection barriers between the solution and the semiconductor. The lines show no-barrier (red line), 0.1 eV (orange line), 0.2 eV (green line), and 0.3 eV (cyan line). The red circle shows the point at which the densities in (a) were simulated. Note that the ion injection barrier shifts the curves (as a $V_T$ shift).

Table 2. Device and material parameters used in the simulations

| Device | Chanel length | 50 $\mu m$ |
|---|---|---|
| | Chanel width | 50 $\mu m$ |
| Semiconductor | Thickness | 100 nm |
| | Hole mobility | 5 cm$^2$/Vs |
| | Source-drain voltage | 0.1 V |
| | Anion diffusivity | $10^{-8}$ cm²/s |
| | Cation diffusivity | ---- |
| Solution | Anion diffusivity | $10^{-6}$ cm²/s |
| | Cation diffusivity | $10^{-6}$ cm²/s |
| | Salt concentration | $10^{20}$cm⁻³ (0.17 M) |

In these simulations, the cations are insoluble in the semiconductor, and the anions injection barrier at the solution/semiconductor interface was implemented by enforcing a ratio of $exp(-E_b/k_BT)$ between the anions on the semiconductor and solution sides of the interface. We first simulated the output characteristics of OECTs with different ionic injection barriers. Fig. 13b shows the output IV curves of OECTs having ionic injection barrier ($E_b$) of 0 eV (red line), 0.1 eV (orange line), 0.2 eV (green line), and 0.3 eV (cyan line). As expected, an injection barrier shifts the curve by the amount equivalent to the barrier energy converted to volts. (denoted $V_T = E_b/q$)

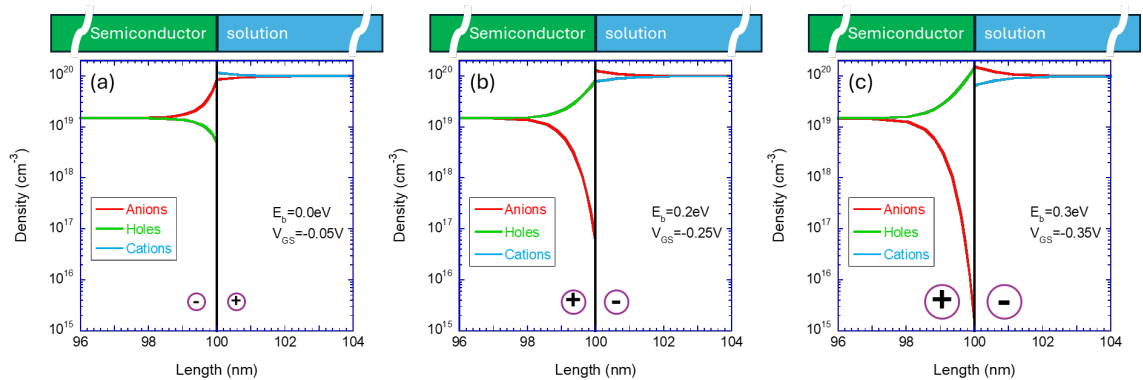

Fig. 14. Simulated charge carriers' density distribution close to the interface (black line) between the semiconductor (left) and the solution (right). The densities are along a cutline in the middle between the source and drain electrodes. The charge carriers are anions (red line), cations (cyan line), and holes (green line). No ionic injection barrier (a) ($E_b$=0), (b) $E_b$=0.2eV and (c) $E_b$=0.3eV. The circled + and − denote the net charge at the interface.

To understand the effect of the anions injection barrier on the interface properties, we plot in Fig. 14 the charge carriers' density distribution close to the interface (black line) between the semiconductor (left) and the solution (right). The charge carriers are anions (red line), cations (cyan line), and holes (green line). To place the devices with different injection barriers on equal footings, we chose gate voltages such that ($V_{GS}$ +$V_T$) is equal for the devices (=50 mV). Indeed, the hole and anion density in the bulk of the devices are identical ($1.5×10^{19}$cm⁻³). The inherent discontinuity at the solution/semiconductor interface is expected to induce some polarization. Fig. 14a shows that without an ionic injection barrier, it extends about 1 nm from the interface. Considering Figures 7b and c, an ionic barrier produces a lower anion density on the semiconductor side, resulting in a higher injection resistance. Also, the ionic barrier induces hole and anion accumulation at the interface, possibly considered a double-layer capacitance. Between $E_b$=0.2eV and $E_b$=0.3eV, the anion density at the interface is reduced by 30, and the integrated net charge close to the interface goes up by a factor of 1.7.

To simulate the transient response of the different devices, we applied a step to the gate bias, ensuring that the ON state corresponds to the same bias used for Fig. 14 ($V_{GS} + V_T$ =50mV). A barrier of 0.1eV had a negligible effect, and Fig. 8 shows the responses for ionic injection barrier ($E_b$) of 0eV (red line), 0.2eV (green line), and 0.3eV (cyan line). The longer time constant associated with a higher barrier aligns with our observation (Fig. 7) that the barrier induces a capacitance and a resistance. Based on our analysis of Fig. 14, the interface resistance is the dominant factor.

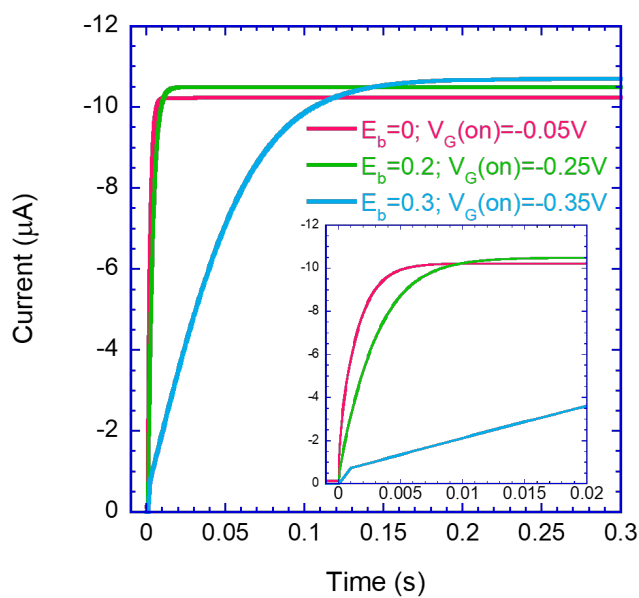

Fig. 15. Transient responses of the devices having different injection barriers following a gate voltage step from 0.2V to the voltage corresponding to $I_D$=10μA (i.e., to the same $V_G$-$V_T$). $V_{DS}$=-0.1V. The inset is a zoom on the first 20 ms.

## Conclusion

We established a general theory of the current transient of ionic-electronic transistors. The dominant effect when the gate voltage is changed is a transient charging of the channel by diffusion of ions. The additional influence of electrolyte capacitance and resistance splits the fundamental time constant of diffusion $\tau_d$ into two different components, $\tau_1$ and $\tau_2$. There are different possible origins to the widely observed transient currents in OECTs. The diffusion charging must be there, as this is the fundamental physical component: the charging of the chemical capacitance. The current transient, however, may be dominated either by the diffusion elements or by external elements. The transient may become much larger than by diffusion alone by impediments of ion transport.

In the BM model, external elements have the leading role. We have provided a more complete classification of the physical factors governing the transient response. The main property determining the shape of the transient is the signs of $\alpha_1$ and $\alpha_2$, which produces capacitive-like decay in the negative case, and inductive-like current rise in the positive cases. The modulation of response shapes according to the intrinsic parameters provides an essential control of synaptic properties as such as STD, STP and LTP in neuromorphic applications. In practice, a combination of experimental techniques is necessary to discriminate the physical origin of the observed transients. The analysis of simulations enables to include more complete effects as the distribution of oppositely charged ions, but confirms the general trends obtained in the theory model. This method provides a convenient framework for the characterization of complex time domains responses of organic electrochemical transistors.

**Acknowledgments**

The work of Juan Bisquert was funded by the Ministerio de Ciencia, Innovación y Universidades (MCIN) and the Agencia Estatal de Investigación (AEI) under grant number PID2022-141850OB-C21 (TAROT), which is co-financed by the European Regional Development Fund of the European Union (MCIN/AEI/10.13039/501100011033/FEDER, EU). Nir Tessler acknowledges the support by the Ministry of Innovation, Science and Technology Israel, and the M-ERANET grant PHANTASTIC Call 2021.

**Associated content**

Data Availability Statement

The data presented here can be accessed at https://doi.org/10.5281/zenodo.13340763 (Zenodo) under the license CC-BY-4.0 (Creative Commons Attribution-ShareAlike 4.0 International).